\documentclass[final,5p,times,twocolumn]{elsarticle}
\usepackage{hyperref}\usepackage{lineno}
\usepackage{multirow}
\usepackage{multicol}
\usepackage{gensymb}
\usepackage{amssymb}
\usepackage{tablefootnote}
\usepackage{caption}
\usepackage{subcaption}
\usepackage{wrapfig}
\usepackage{amsmath}
\usepackage[dvipsnames]{xcolor}

\usepackage{dblfloatfix}
\usepackage{fixltx2e}
\usepackage{placeins}
\newcommand{\micronsq}{$\mu$m$^2$~}
\newcommand{\microns}{$\mu$m~}
\newcommand{\micron}{$\mu$m}
\newcommand{\nq[1]}{$\cdot 10^{#1} \; n_{eq}/cm^2 $}
\begin{document}

\begin{frontmatter}

\title{4D Tracking: Present Status and Perspective}

%% Authors, affiliations in footnotes:
\author[1]{N. Cartiglia\corref{corrauth}}
\cortext[corrauth]{Corresponding author}
\ead{cartiglia@to.infn.it}
\author[1,2]{R.~Arcidiacono}
\author[1,3]{M.~Costa}
\author[1,2]{M.~Ferrero}
\author[3]{G.~Gioachin}
\author[1]{M.~Mandurrino}
\author[1]{L.~Menzio}
\author[1]{F.~Siviero}
\author[1,3]{V.~Sola}
\author[1,3]{M.~Tornago}

\address[1]{INFN, Torino, Italy}
\address[2]{Universit\`a del Piemonte Orientale, Italy}
\address[3]{Universit\`a di Torino, Torino, Italy}

\begin{abstract}
The past ten years have seen the advent of silicon-based precise timing detectors for charged particle tracking. The underlying reason for this evolution is a design innovation: the Low-Gain Avalanche Diode (LGAD). In its simplicity, the LGAD design is an obvious step with momentous consequences: low gain leads to large signals maintaining sensors stability and low noise, allowing sensor segmentation. Albeit introduced for a different reason, to compensate for charge trapping in irradiated silicon sensors, LGAD found fertile ground in the design of silicon-based timing detectors. Spurred by this design innovation, solid-state-based timing detectors for charged particles are going through an intense phase of R\&D, and hybrid and monolithic sensors, with or without internal gain, are being explored. This contribution offers a review of this booming field.
\end{abstract}


\end{frontmatter}

%\clearpage

%\linenumbers

\section{Introduction}
In many of the past and present high-energy physics experiments, the temporal information of charged particles is mainly used to perform particle identification via Time-of-Flight (ToF). ToF systems are usually quite large as they require long distances between production and detection points. For this reason, they use low-granularity large-area media such as scintillators, gaseous detectors, or Cherenkov-based detectors (MCP arrays or DIRC). An up-to-date review of ToF systems and associated R\&D programs is presented in ~\cite{Detector:2784893}. The temporal performances of solid state-based trackers were generally insufficient for accurate ToF systems, mostly due to the low amplitude of the signal generated by an impinging particle and the short flight distance. An early proposal to build a silicon-based ToF system is reported in ~\cite{5733383}, using fast-shaping electronics and 3D sensors. The proposed possible application for such a detector is a small-angle, far-forward detector at colliders where the relative timing of the two scattered particles could locate their vertex position among several possible vertices. 

In the past few years, the situation has radically changed, mostly due to the introduction of the Low-Gain Avalanche Diode 
 (LGAD)~\cite{FERNANDEZMARTINEZ201198, LGAD1} design (introduced to compensate for the loss of signal due to charge trapping in irradiated sensors) and its subsequent optimization for timing application in Ultra-Fast Silicon Detector (UFSD)~\cite{CARTIGLIA2015141}. This R\&D has spurred a strong evolution in the field of accurate timing using silicon detectors, now including sensors with and without internal gain in hybrid or monolithic architectures. These approaches aim to reach excellent temporal precision by optimizing different aspects of the detector chain, such as using larger signals, lower noise, or low detector capacitance. 
 
 This renewed interest for trackers able to perform the concurrent measurements of the spatial and temporal coordinates (the so-called 4D tracking) is due to the combination of technological advances (LGAD, use of SiGe, HVCMOS) with the needs of future experiments, where 4D tracking is an essential tool to reach the physics goals~\cite{Detector:2784893}. A summary of the critical R\&D paths in 4D tracking is given in the list of Detector R\&D Themes (DRDTs) reported in ~\cite{Detector:2784893}:
\begin{itemize} 
\item Understand the ultimate limit of precision timing in sensors with and without internal multiplication; 
\item Develop sensors with internal multiplication with 100\% fill factors and pixel-like pitch; 
\item Investigate production of sensors with internal multiplication in a monolithic design; 
\item Increase radiation resistance, push the limit of 3D sensors and explore LGAD and MAPS capabilities; 
\item Investigate the use of BiCMOS MAPS, exploiting the properties of SiGe. 
\end{itemize} 

\begin{figure}[htb]
\begin{center}
\includegraphics[width=0.5\textwidth]{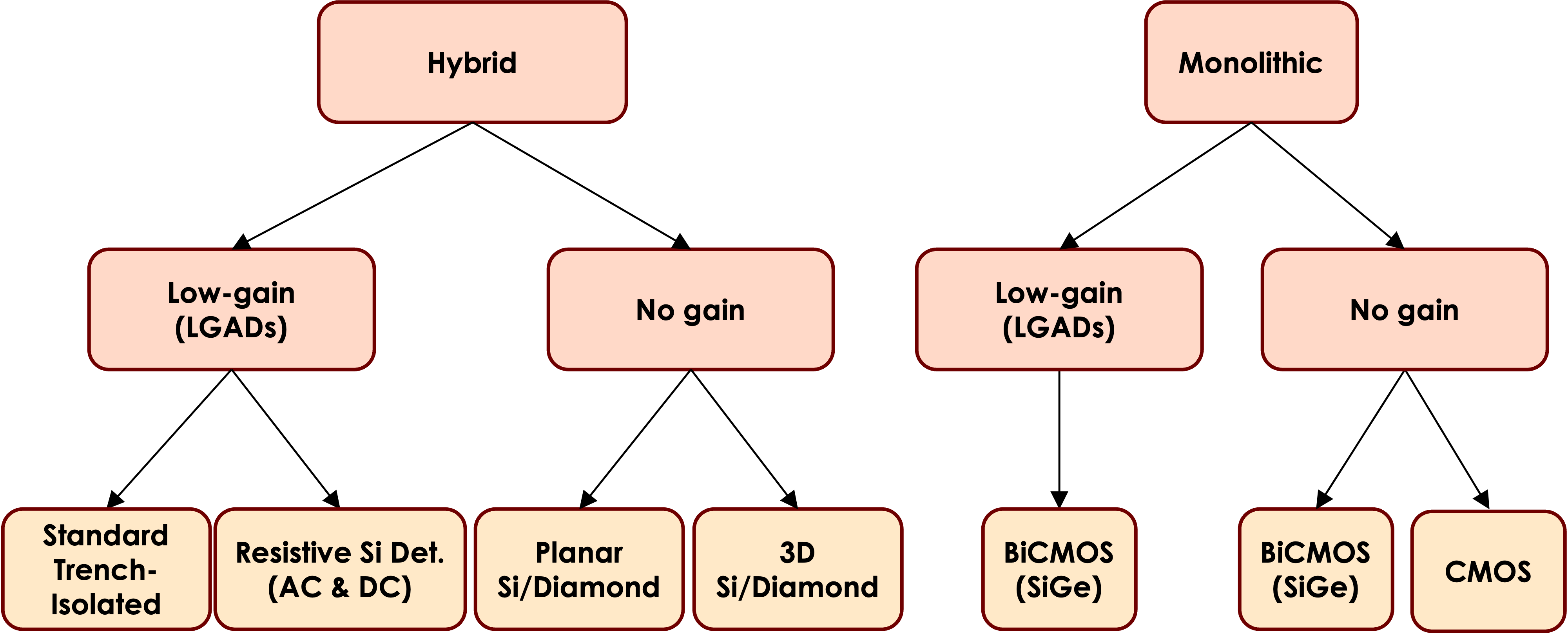}
\caption{R\&D activities in sensors for 4D tracking}
\label{fig:det}
\end{center}
\end{figure}

In order to cover the above points systematically, in this report, the sensors are divided into 4 broad families: hybrid and monolithic, with and without gain. This approach is shown in Figure~\ref{fig:det} where, for each family, the most relevant designs or technologies are reported. 

\section{Signal formation and time-tagging: aide-memoire}
\label{sec:sig}

The shape of the induced current signal can be calculated using the Ramo-Shockley theorem ~\cite{Shockley, Ramo}. This theorem states that the current induced by a charge carrier is proportional to its electric charge $q$, the drift velocity $v$, and the weighting field $E_w$:
\begin{equation}
\label{eq:ramo}
i = q v E_w.
\end{equation}
Variations in the shape of the current $i$ directly impact the timing performances since, in most systems, the time of a hit is set when the signal reaches a certain value $V_{Th}$. Equation~\ref{eq:ramo} therefore indicates that a constant drift velocity and a constant weighting field are necessary conditions to reach good timing performances. Of all possible electrode geometries, the one that achieves these two conditions is that of the parallel plate capacitor. For a detector, this requirement is translated into having the implant width as large as the pixel pitch and the pixel pitch to be several times larger than the sensor thickness. Detectors whose geometries significantly differ from a parallel plate capacitor have degraded performances. 

 In very general terms, the time resolution $\sigma_t$ of a detector can be expressed as the sum of several terms: (i) jitter, (ii) fluctuations of the ionisation process producing shape and amplitude variations, (iii) signal distortion, and (iv) TDC binning:
\begin{equation}
\label{eq:main}
\sigma_t^2 = \sigma_{Jitter}^2 + (\sigma_{Landau\; Noise}+\sigma_{Total\; ionization})^2 + \sigma_{Distortion}^2 + \sigma^2_{TDC}.
\end{equation}

Let's analyze the terms (see~\cite{ROPP} for details) of Eq.~\ref{eq:main}: 
\begin{itemize}
\item $\sigma_{Jitter} = N/(dV/dt)$: due to the effect of the noise $N$ when the signal is approaching $V_{Th}$ with a $dV/dt$ slope. $\sigma_{Jitter}$ is where the contribution from the electronics is apparent since $N$ is dominated by the electronic noise. 
\item $\sigma_{Total\; ionization}$: due to the fact that the energy deposited by a MIP follows a Landau distribution. Signals with different amplitudes cross a fixed threshold at different times (the so-called time walk effect). $\sigma_{Total\; ionization}$ is minimized by an appropriate electronic circuit (either Constant Fraction Discriminator or Time over Threshold). 
\item $\sigma_{Landau\; Noise}$: due to signal shape variation on an event-by-event basis caused by the random nature of electron-hole pairs creation along the particle path. $\sigma_{Landau\; Noise}$ is absent in 3D detectors, and it is minimized in sensors with parallel plate geometry, while it is enhanced by internal gain. 
\item $\sigma_{Landau\; Noise}\times \sigma_{Total\; ionization}$: due to the correlation between large signals and non-uniform ionization. The events in the high tail of the Landau are mostly due to the presence of localized clusters of ionization. Given that the ionization is very non-uniform for these events, their temporal resolution is worse than that of signals with an amplitude around the Landau most probable value ~\cite{FSiv}.
\item The $\sigma_{Distortion}$: due to signal shape variations as a function of the hit position in the pixel. Two factors contribute: (i) non-uniform weighting field and (ii) non-saturated drift velocity. Both terms are reduced to be sub-leading contributions by using a "parallel plate geometry" and operating the sensor at a bias voltage where the velocity of the carriers is saturated.  
\item $\sigma_{TDC}$: due to the TDC digitization binning.
\end{itemize}

\section{Timing layers and 4D tracking}
\label{sec:TL}
\begin{figure}[h]
\begin{center}
\includegraphics[width=0.5\textwidth]{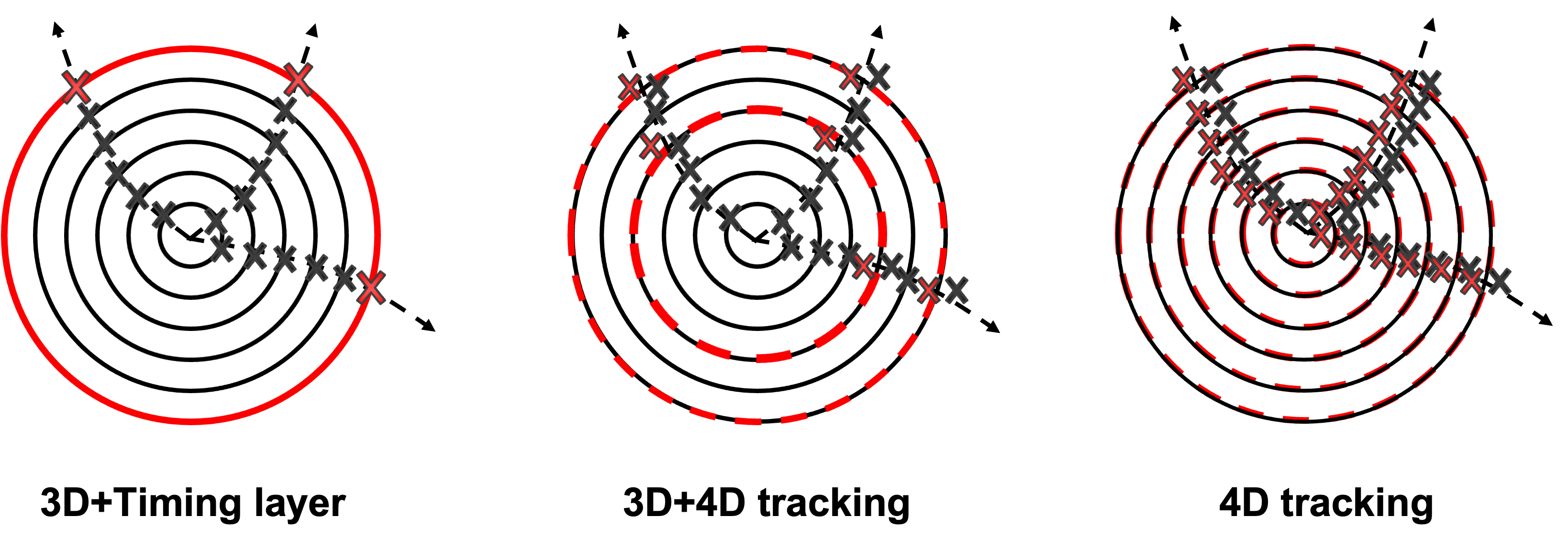}
\caption{Possible implementations of a tracker system with timing capability. }
\label{fig:4DTracking}
\end{center}
\end{figure}

Present tracking systems in high-energy physics experiments are complex, with millions of separated pixels, state-of-the-art mechanical and cooling systems, and massive data transmission~\cite{Hartmann}. These systems are optimized for best tracking performances, and the pixels are sized to achieve the single-point resolution needed by the specific application. Including the temporal coordinate in such systems is a formidable task that requires a complete redesign. The request for timing information implies having the space to place the front-end electronics, the cooling power to remove the extra heat generated by the timing circuitry, the distribution of a reference clock, and the data transmission capabilities to send off-detectors the additional bits. To make the design of 4D tracking systems even more difficult, the request to have timing information is often coupled with the requests for excellent spatial resolution, for example, less than 10 $\mu$m, and a very low material budget. A compilation of future requests can be found in Figure 3.3 of ~\cite{Detector:2784893}. The path to developing a full 4D tracking system is very challenging, and it will be accomplished via a series of incremental steps. The ATLAS~\cite{ATLAS_HGTD} and CMS~\cite{CMS_MIP} collaborations, in their respective upgrades for HILUMI-LHC, have taken the first step: the addition of a timing layer to their 3D tracking systems. In this design, sketched in the left pane of Figure~\ref{fig:4DTracking}, the timing coordinates of the tracks (red crosses) are determined by a dedicated layer that does not also provide the spatial coordinates (black crosses). 

A foreseeable evolution of this initial step is the configuration shown in the center pane of Figure~\ref{fig:4DTracking}: a tracker with mostly 3D layers with a few 4D layers.
The number of 4D layers depends on the overall needed track temporal precision: albeit multiple 4D layers look more complex than a single one, in these systems, the single point precision might be relaxed, simplifying the design and reducing power consumption. The right pane of Figure~\ref{fig:4DTracking} shows the complete 4D tracking configuration, where each layer contributes to both the spatial and temporal coordinates. For such systems, also the reconstruction process benefits from the temporal information, enforcing the constraint of time of flight compatibility between layers and reducing the overall number of possible hit combinations. 

It is worth stressing that not all experimental environments require a total redesign of their tracking systems. In most applications, a timing layer is enough to assure most of the benefits brought about by the timing information, and when the required temporal precision is somewhat relaxed (above 75 - 100 ps), the increase in power consumption is limited.

\section{Interplay among position and temporal resolutions, occupancy, material budget, and power}
The interplay among position and temporal resolutions, occupancy, material budget, and power are fairly intricate. In the following, a series of points highlight some of these dependencies and provide information on present and future systems.
\begin{itemize} 
\item In present tracking systems, the pixel size is determined either by the need to achieve a given position resolution or by the need to keep the occupancy below a given number, for example, 1-3\%. As a general rule, occupancy determines the pixel size in the innermost layers of a tracking system, while position resolution determines the pixel size in the outer layers (which is by far the largest fraction of the total tracker area). 
\item Presently, the ATLAS, CMS, and LHCb silicon trackers generate about 0.5- 1 W/cm$^2$. 
\item Trackers that have small pixels in their outer layers are not power-efficient as, at any given moment, they have millions of idling electronics channels.
\item The power needs are driven by the density of electronic channels and not by the silicon volume, so it is higher for smaller pixel sizes. For example, the power consumption of the ALICE MAPS-based tracker is about 300 mW/cm$^2$ in the inner layers and 100 mW/cm$^2$ in the outer layers. 
\item The highest cooling power is required by the inner layers at hadron trackers, which combine very high particle density (i.e. need for small pixels) and radiation damage (need to keep the silicon bulk as cold as possible). 
\item At future $e^+e^-$ colliders such as CLIC or FCC-ee, the material budget constraints are so severe (about the equivalent of 100 $\mu$m of silicon per layer) to require air (or Helium) cooling, limiting power consumption to about 100 mW/cm$^2$.
\item In 4D-trackers, the power increases significantly as the circuitry for timing determination uses more power than that for position. This increase depends on the performances required, more power for better precision, and the number of pixels, many pixels covering the same area use more power than a single pixel covering the same area.  
\item An important benefit of the LGAD technology is its power efficiency: internal multiplication requires almost no power, decreasing the amount of power needed by the electronics. The ATLAS and CMS timing layers are an example of how the combination of the LGAD design with large pixels, $\sim$60 channels/cm$^2$, generates the same power consumption of a pixel system with $\sim$10k channels/cm$^2$, about 0.3-0.5 W/cm$^2$. 
\item 4D trackers need pixels large enough to allow space for the electronics and to limit power consumption. From a power point f view, the present design of most outer trackers layers, millions of small pixels idling, cannot be duplicated in 4D trackers. Ideally, the pixel size in 4D tracking has to be determined by occupancy, strongly reducing the number of pixels. 
\item The use of charge sharing among nearby pixels allows using larger pixels since, in this design, the position resolution is much better than pixel size/$\sqrt(12)$. 
\item In present trackers, charge sharing is based on Lorentz-angle drift and requires thick sensors; for this reason, it cannot be used in systems with a limited material budget, such as those at future lepton colliders. 
\item In timing circuits, the power consumption does not decrease significantly with the technological node of the electronics (130 nm, 65 nm, or 28 nm). 
\end{itemize}
Table~\ref{tab:power} reports a compilation of front-end ASICs and monolithic systems. The first five systems use a hybrid design; the bottom 4 are monolithic. The first four systems are very advanced or completed, while the bottom 5 are in their R\&D phase, so their performances might change rapidly. 
\begin{table*}[htb]
 \centering
 \tabcolsep3pt
 \begin{tabular}{cccccc}
 \hline
  Name & Sensor& Node & Pixel size& Temporal & Power\\
   & & [nm] & [$\mu$m$^2$]& precision [ps] & [W/cm$^2$]\\	\hline\hline
   ETROC & LGAD & 65 & 1300x1300 & $\sim$ 40 & 0.3 \\	\hline
ALTIROC & LGAD & 130 & 1300x1300& $\sim$ 40 & 0.4\\	\hline
TDCpix & PIN & 130 & 300x300 & $\sim$ 120 & 0.32 matrix + 4.8 periphery \\ \hline
TIMEPIX4 & PIN, 3D & 65 & 55x55 & $\sim$ 200 & 0.4 analog + 0.3 digital\\	\hline
TimeSpot1 & 3D & 28 & 55x55 & $\sim$ 30 ps & 3-5\\	\hline
FASTPIX & MAPS & 180 & 20x20 & $\sim$ 130 & 5-10\\	\hline
miniCACTUS & MAPS & 150 & 500x1000 & $\sim$ 90 & 0.15 – 0.3\\	\hline
MonPicoAD & MAPS & 130 SiGe & 100x100 & $\sim$ 36 & 1.8\\	\hline
Monolith & Multi Junct. MAPS & 130 SiGe & 100x100 & $\sim$ 25 & 0.9\\		\hline
    \end{tabular}
 \caption{Compilation of front-end ASICs and monolithic systems. The first 5 systems use an hybrid design, the bottom 4 are monolithic. The first 4 systems are very advanced or completed, while the bottom 5 are in their R\&D phase, so performances might change rapidly. }
 \label{tab:power}
\end{table*}

\section{Sensors without internal gain}
\begin{figure}[h]
\begin{center}
\includegraphics[width=0.5\textwidth]{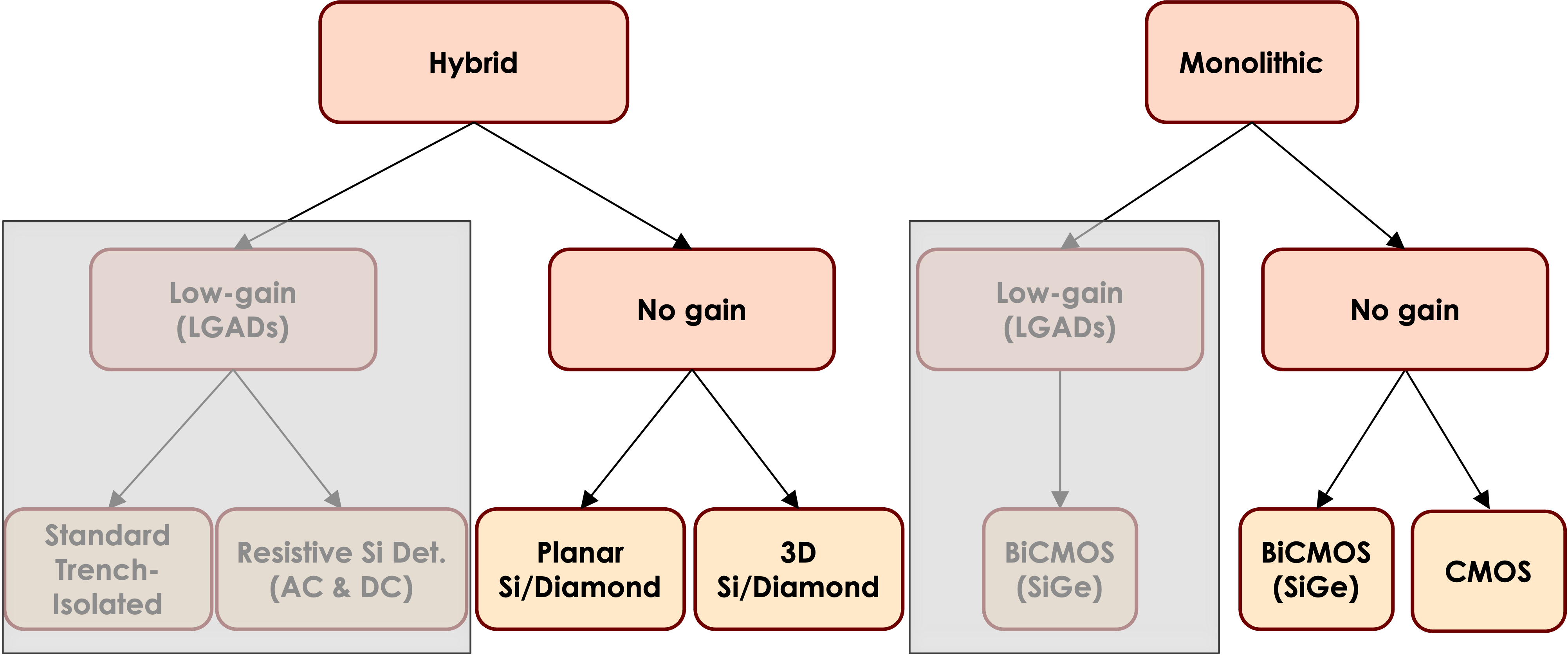}
\caption{ Sensors for 4D tracking without internal gain}
\label{fig:sensng}
\end{center}
\end{figure}

The sensors presented in this section are shown in Figure~\ref{fig:sensng}: hybrid sensors with planar and 3D geometry, and monolithic sensors.

\begin{figure}[htb]
\centering
\includegraphics[width=0.5\textwidth]{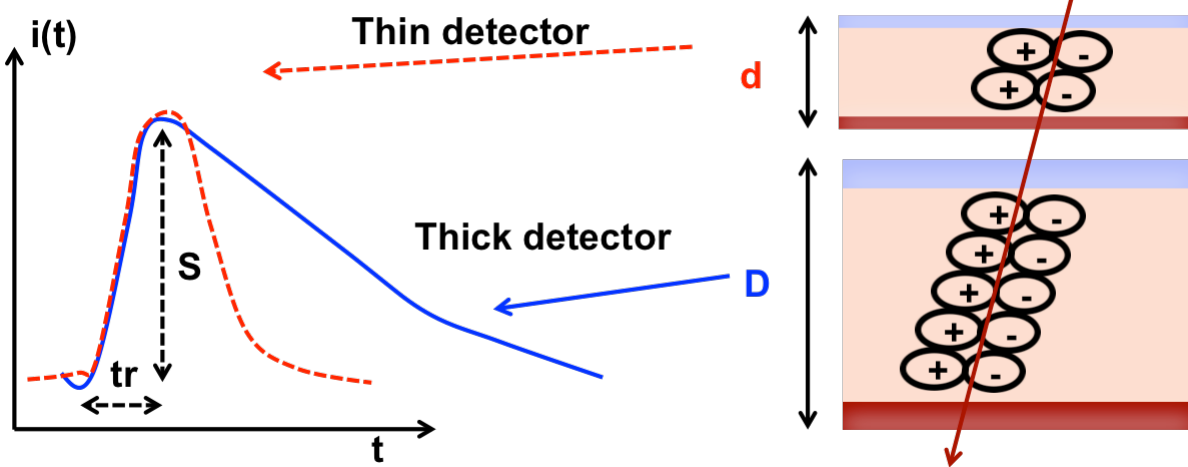}
\caption{The initial signal amplitude in planar sensors does not depend on their thickness: thin and thick detectors have the same maximum current, while thick detectors have longer signals. The rise time of the signal, $t_r$, is due solely to the read-out electronics.}
\label{fig:cu}
\end{figure}

The temporal precision of planar silicon (and diamond) sensors without internal gain is limited by the smallness of the signal amplitude, about 3 (1.7) fC in a 300 $\mu$m thick silicon (diamond) sensor. Interestingly, the peak signal current does not depend on the sensor thickness~\cite{CARTIGLIA2015141}: thick sensors have a larger number of initial e/h pairs; however, each pair generates a lower initial current since the weighting field is inversely proportional to the sensor thickness, Figure~\ref{fig:cu}. This interplay is such that the MPV peak current in planar sensors is always the same; in silicon, it is about  $I_{max} \sim 1-2$ $\mu$A. In 3D geometry, this fact is not true, and the maximum current is directly proportional to the sensor thickness. 
 
The rise time of the signal, $t_r$, is due solely to the read-out electronics as the intrinsic rise time of the signal is that of the passage of the particle. Under these circumstances, the front-end electronics slew rate and noise are the factors dominating the temporal resolution (assuming no distortions due to the carriers velocity and weighting field).

\subsection{Monolithic systems}
\label{sec:ngmono}

Monolithic systems have been developed both in CMOS and SiGe technology.  
In the following, 2 examples of CMOS monolithic sensors are presented, FASTpix, geared at very small pixels, and miniCactus for hard radiation environments and large pixels. 
FASTpix {~\cite{instruments6010013} is designed in a modified 180 nm CMOS imaging device technology, with small, low-femtofarad collection electrodes on high-resistivity sensing layers. The FASTpix demonstrator consists of 32 mini matrices with hexagonal pixels, split into four groups with a pixel pitch of 8.66 \micron, 10 \micron, 15 \micron, and 20 \micron. The defining feature of this project is that an innovative doping implant is shaping the electric field to uniformize the drift path within a pixel cell. Beamtest results showed that this architecture is able to obtain a temporal resolution of about 120-180 ps. 

MiniCactus has been designed in the LFoundry 150 nm HV-CMOS process with the objective of developing a radiation-hard monolithic timing sensor using a commercial HV-CMOS process. In this design, the bulk is fully depleted (either 100 \microns or 200 \micron), and charge collection happens by drift and not diffusion, so it is suited for timing applications. MiniCactus has reported a temporal precision of about 90 ps ~\cite{miniCactus}, obtained in a recent beamtest with a pixel size of 0.5x1 mm$^2$. 

The last example of a monolithic detector without internal gain, the MonPicoAD ~\cite{Iacobucci_2022}, differs from the previous two in the choice of technology: it uses SiGe instead of CMOS. The monolithic silicon pixel detector prototype has been produced in the SiGe BiCMOS SG13G2 130 nm node technology by IHP. The ASIC contains a matrix of hexagonal pixels with a pitch of approximately 100 \micron. The choice of the SiGe technology allows for a faster slew rate and lower noise, reducing the jitter term. This technology, combined with small input capacitance, allowed the MonPicoAD to obtain a resolution of about 35 ps in a recent beamtest. This result is obtained at the highest pre-amplifier current, yielding a power consumption of about 40 W/cm$^2$.

\subsection{Hybrid systems}
This group includes silicon and diamond 3D sensors with columns or trenches and planar sensors.

\subsubsection{3D sensors, silicon, and diamond}
 
3D silicon sensors are well known for their radiation resistance, and they are currently used successfully in the ATLAS inner pixel layer ~\cite{Capeans:1291633}. 3D sensors are good candidates as timing sensors since the drift time is very short~\cite{3Dpaper}. Remarkably, the current signal generated by a particle in a sensor with 3D geometry does not suffer from local ionization fluctuation, $\sigma_{Landau\; Noise}$, since the induction mechanism happens perpendicularly to the charge distribution generated by the impinging MIP. In the standard 3D implementation, on the left side of Figure~\ref{fig:3D}, both electric and weighting fields change rapidly, yielding a position-dependent signal shape that degrades the temporal resolution. However, since the signal in 3D sensors is very short given the small pixel size, even with this less-than-ideal geometry, a temporal resolutiopeakn of about 30 ps has been achieved~\cite{KRAMBERGER201926}. In recent years, the design of the 3D sensors has been modified by the TimeSpot project replacing the columns with trenches to achieve more uniform electric and weighting fields, left side of Figure~\ref{fig:3D}~\cite{LAI2020164491}. The present realization of the 3D trench detector has a pixel size of 50 $\mu$m; it is 200 $\mu$m thick, and the operating voltage is about 200V when not irradiated. For a new detector, the MPV signal charge is 2.2 fC, while it is reduced to 1 fC for the irradiated case due to charge trapping. The 3D trench layout leads to an almost ideal sensor since it combines very short drift time, uniform fields, and the absence of the $\sigma_{Landau\; Noise}$ contribution. The intrinsic time resolution of 3D trench detectors has not been established yet since, up to now, the performances of the electronics have limited it: a value in the range 10-15 ps has been suggested. The TimeSpot ASIC, designed in 28 nm technology, is tailored to the readout of trenched detectors with a 50 $\mu$m pitch and presently reaches a temporal resolution of about 30 ps with a power budget of 2-3 W/cm$^2$. The 3D geometry is also pursued in the design of diamond detectors for timing applications. The main limiting parameter is presently the electrode resistance, achieving a precision just below 100 ps and efficiency larger than 99\% ~\cite{Anderlini:2021pei}

\begin{figure}[htb]
\begin{center}
\includegraphics[width=0.5\textwidth]{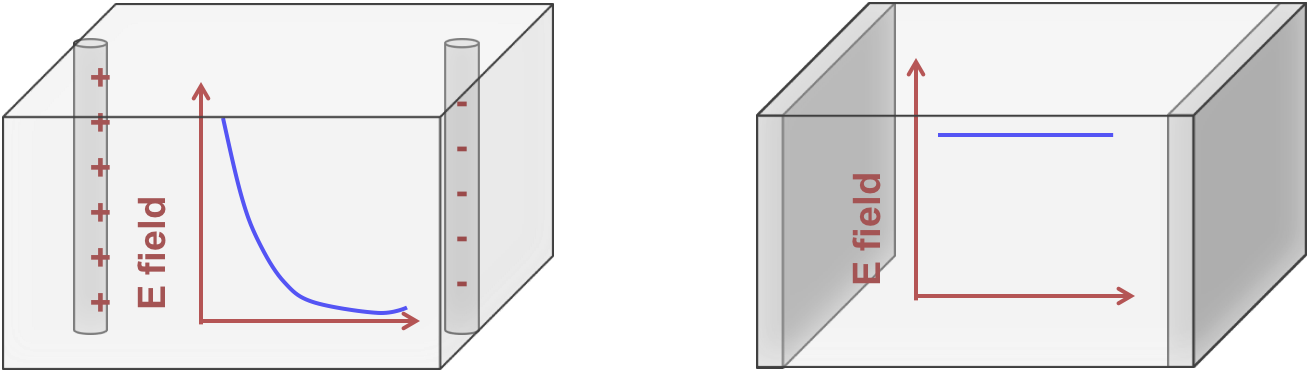}
\caption{Left side: 3D column sensors have electric and weighting fields changing rapidly with the position. Right side: 3D trench sensors have parallel plate-like geometry, with constant fields, ideal for timing measurements.}
\label{fig:3D}
\end{center}
\end{figure}

\subsubsection{Planar sensors, silicon, and diamond}
 
 The most advanced 4D-tracker detector using planar sensors in a hybrid configuration is the NA62 GigaTracker ~\cite{Rinella_2019} (GTK). The GTK comprises three stations, each made of 18000 pixels of 300x300 \micronsq for a total area of 60.8 × 27 mm$^2$. The sensors are 200 \microns thick and are read out by two rows of five ASIC called TDCPix, thinned to 100 \micron. The TDCPix functionalities are the sensor hit signal amplification, discrimination, digitization, time- stamping, and the transmission of the resulting digitized data off the chip. The TDCPix was designed in a commercial CMOS 130 nm technology and achieved a single hit temporal resolution of about 130 ps. This value of resolution is a good indication of the achievable performances of this architecture. 
 
 Another very interesting development in 4D-tracking using a hybrid configuration is provided by the Timepix4 ASIC~\cite{Llopart_2022}. This application aims to achieve an excellent position resolution with good time-tagging capability. 
 Timepix4 is a 24.7 x 30.0 mm$^2$ read-out ASIC consisting of 448 x 512 pixels which can be bump bonded to a sensor with 55 \microns square pixels. Timepix4 reaches a hit resolution of about 200 ps. The analog power consumption depends on the exact biasing conditions used, and in the default configuration, it is estimated to be about 400 mW/cm$^2$. The digital power consumption depends on the clock frequency used and, in data-driven mode, increases depending on the incoming hit rate. Below 3 Mhits/mm$^2$/s and at full clock speed, it is below 200 mW/cm$^2$. 
 
 The PPS detector of the CMS experiment uses planar diamond sensors to tag protons diffracted at very small angles ~\cite{Bossini_2020}. The requirement on radiation hardness has driven the choice of diamond sensors: the sensors have to sustain highly non-uniform irradiation, with a peak of about 5$\cdot$10$^{15}$ protons/cm$^2$ in the near beam region for an integrated LHC luminosity of 100 fb$^{-1}$ (which represent the order of magnitude delivered by LHC). The PPS diamond sensors are made of scCVD crystals with a surface of 4.5x4.5 mm$^2$ and a thickness of 500 \microns with a total active surface coverage of about 20x4.5 mm$^2$. In beamtest, the performance of a tagging station comprising two diamond planes has been measured to be about 50 ps, while during operation at LHC, due to the much harsher conditions, the resolution of a PPS station is about 120 ps. An interesting review on planar diamond sensors can be found in~\cite{10.3389/fphy.2020.00248}.

\section{Sensors with internal gain}
The sensors presented in this section are shown in Figure~\ref{fig:sensg}: hybrid and monolithic sensors based on the LGAD technology
\begin{figure}[h]
\begin{center}
\includegraphics[width=0.5\textwidth]{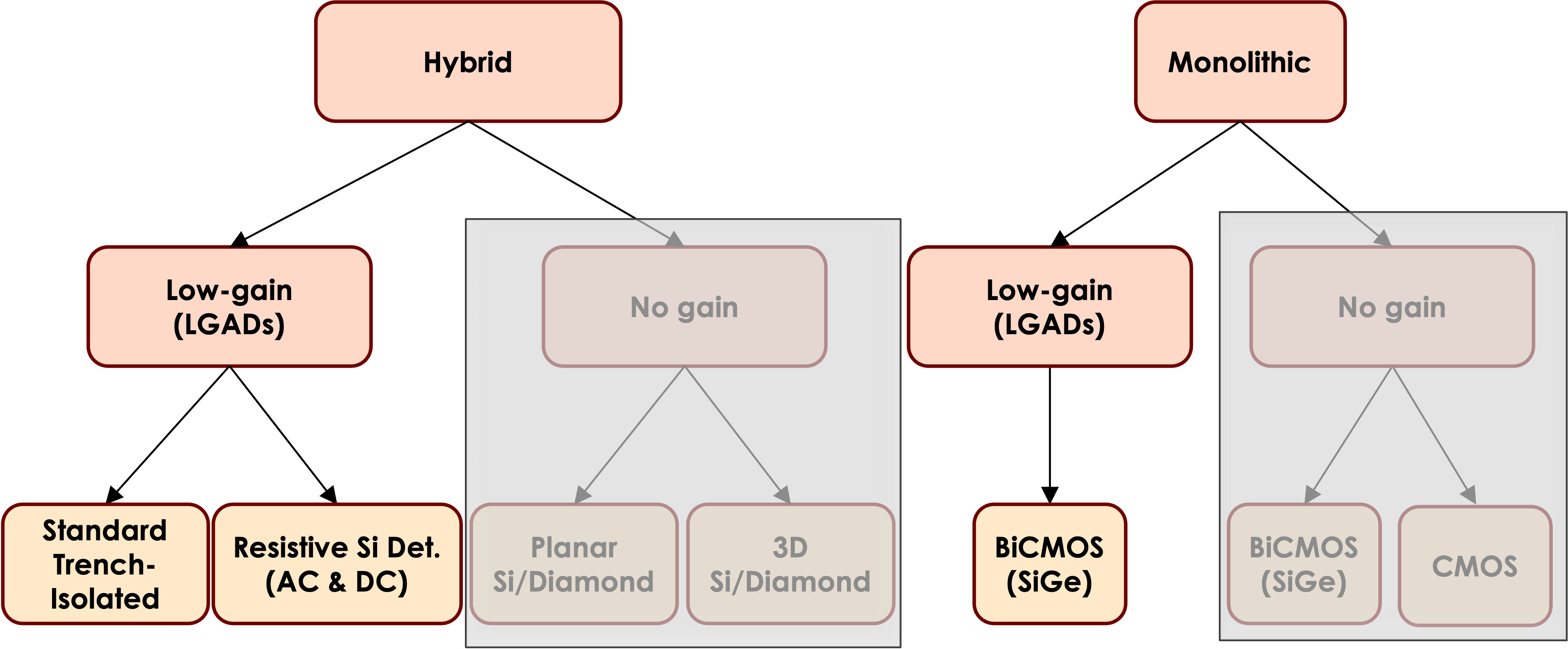}
\caption{ Sensors for 4D tracking with internal gain}. 
\label{fig:sensg}
\end{center}
\end{figure}

The core of the LGAD design~\cite{LGAD1} is an additional implant of doping situated in the proximity of the read-out electrode. Multiplication happens when the electrons (in the standard n-in-p LGADs) enter the high field region generated by this implant. The multiplication process increases the sensor noise more than the signal amplitude (due to the excess noise factor); however, since the total noise is largely dominated by the electronic noise, the overall effect is a strong jitter reduction~\cite{Cartiglia_2016}. In the LGAD design, the effects of non-uniform ionization are enhanced, and the term $\sigma_{Landa\; Noise}$ is often the dominant source of the temporal resolution. Figure~\ref{fig:lgad} shows on the left side the simulation of the ionization pattern of a MIP, performed by the Weightfield2 program~\cite{CENNA201514}, and on the right side a collection of simulated pulses for a 50 \microns thick LGAD with a gain of about 10-15. The shape distortions visible on the signal rising edge limit the achievable resolution, and this intrinsic resolution depends on the sensor thickness~\cite{CARTIGLIA2015141}. The Landau noise in 50 \microns thick sensor is about 30 ps, and it becomes about 25 ps for a thickness of 50 \microns ~\cite{FSiv}.

\begin{figure}[h]
\begin{center}
\includegraphics[width=0.5\textwidth]{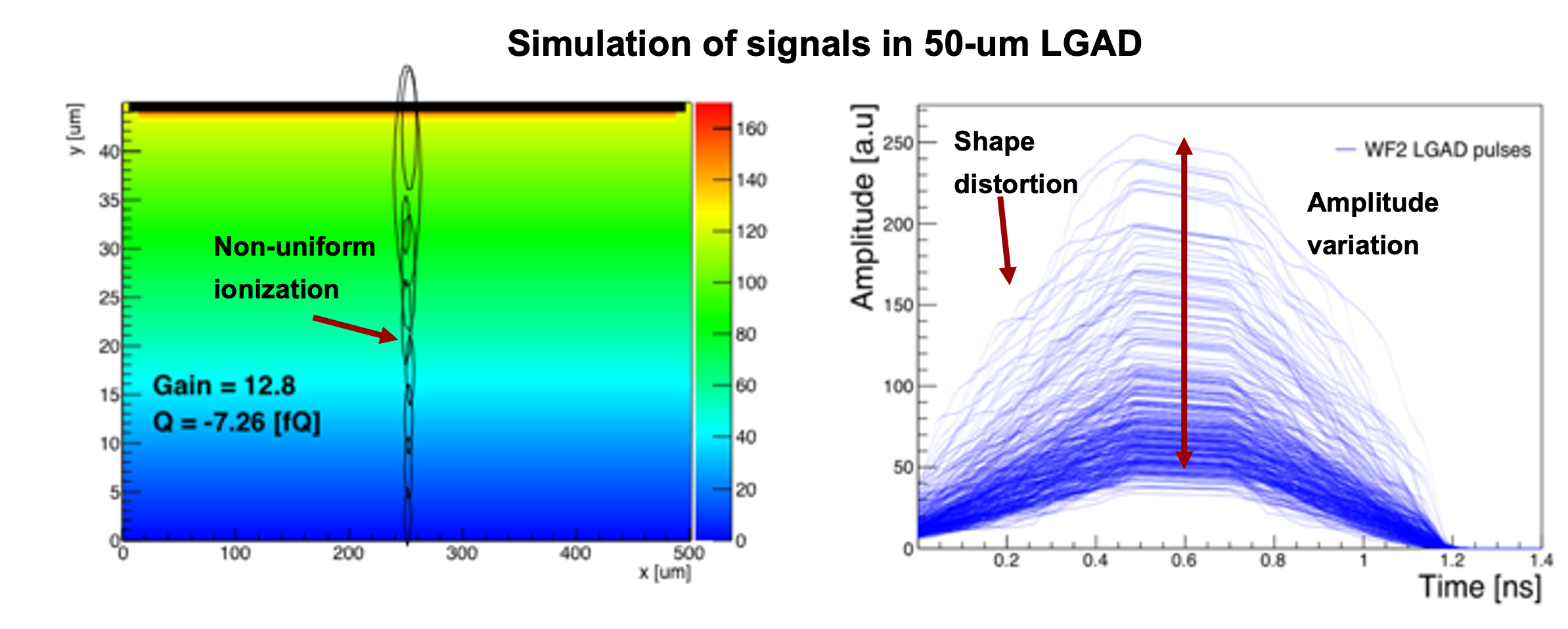}
\caption{Left: Simulation of the energy deposition in an LGAD. Right: signal variations due to the fluctuations of the ionizing process}. 
\label{fig:lgad}
\end{center}
\end{figure}

\subsection{Monolithic systems}
At the moment, the only project that merges the monolithic approach with an internal gain is the Monolith project ~\cite{9620045}. The project merges the already excellent performances obtained by MonPicoAD ASIC, see sec.~\ref{sec:ngmono}, with internal multiplication, where the gain layer is obtained with an additional deep junction. This second pn junction forms a continuous gain layer that operates in avalanche mode. The gain layer is not implanted as commonly done right underneath the pixel, but instead, it is placed a few \microns from the backside of the silicon bulk. In the simulation reported in ~\cite{9620045}, the overall thickness of the sensor is about 5 \microns, and a resolution of 5.8 (3.5) ps is obtained with a gain of about 15 (100). The project recently reported a resolution of about 24 ps as a first result~\cite{9620045,Munker}, with a yet-to-be finalized sensor and electronics. 

\subsection{Standard LGAD and TI-LGAD}

LGAD optimized for timing, the so called Ultra-Fast Silicon Detector~\cite{ROPP, CRC} or simply LGAD, have been the subject of an intense R\&D study in the past few years, and they are now considered a mature enough design to be employed the ATLAS and CMS timing layers~\cite{ATLAS_HGTD} and CMS~\cite{CMS_MIP}. The sensor design used by the two collaborations is shown in Figure~\ref{fig:lgad1}
\begin{figure}[h]
\begin{center}
\includegraphics[width=0.5\textwidth]{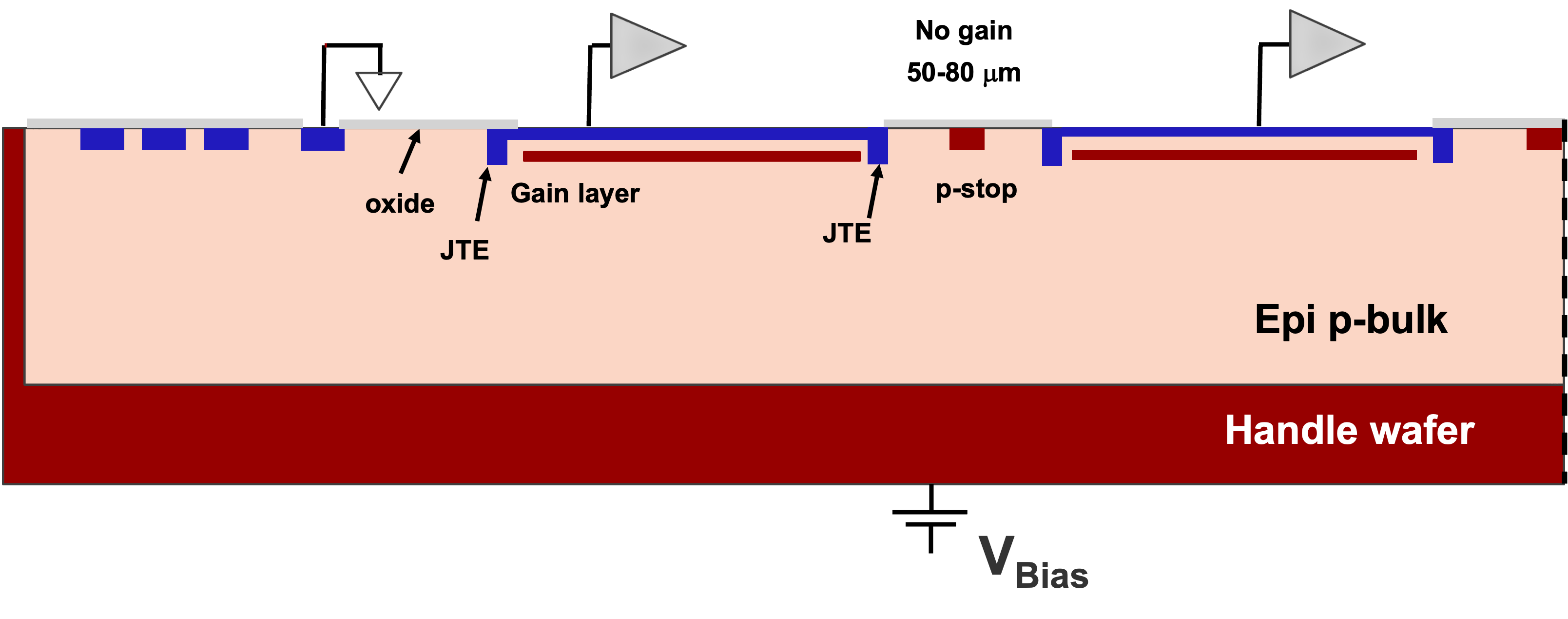}
\caption{Sketch of the LGAD design used in the proposed ATLAS and CMS timing layers}. 
\label{fig:lgad1}
\end{center}
\end{figure}

The most relevant aspects of the design are:
\begin{itemize}
\item ATLAS (CMS) 15x15 (16x16) pads, each 1.3 x 1.3 mm$^2$ 
\item Active thickness 45 - 55 \microns
\item Gain when new 20-30
\item Radiation resistant up to 1-2\nq[15] with carbon infusion in the gain layer
\item Interpad no-gain distance 50-80 \microns
\item 100\% efficiency
\item $\sigma_{Landa\;Noise} \sim$ 30 ps.
\end{itemize}
Both collaborations are developing new read-out ASICs for their respective timing layers: ALTIROC ~\cite{9507972} designed in 130 nm CMOS technology by ATLAS and ETROC~\cite{osti_1623361} in 65 nm CMS technology by CMS. ALTIROC and ETROC are the first attempts to develop large ASICs (about 2x2 cm$^2$) dedicated to reading LGAD sensors, aiming at a sensor-electronics combined single hit resolution below 50 ps. Both ASICs use a preamplifier-discriminator front-end that generates the digital pulse, which provides the Time-Of-Arrival (TOA, leading-edge) and Time-Over-Threshold (TOT, pulse width) information needed for the time walk corrections. The power consumption is between 0.4 and 0.5 W/cm$^2$ for both systems.

One important drawback of the ATLAS and CMS sensor design is the no-gain distance of about 60-80 \microns between two adjacent pads. Given the large pixel size in the ATLAS and CMS timing layers, this feature yields a fill-factor reduction of about 10\% (somewhat compensated by the use of multiple detection layers); however, such considerable no-gain distance makes the use of this design impractical for small pitch sizes. 
A very promising solution to this problem is the introduction of shallow trenches to replace the JTE and p-stop~\cite{ARCIDIACONO2020164375, 9081916}, the so-called TI-LGAD. The introduction of trenches~\cite{Bishit} lowers the no-gain distance to 0-10 \microns depending on the specifics of the implementation while maintaining complete pad isolation. Extensive testing of the first FBK (Fondazione Bruno Kessler) TI-LGAD production ~\cite{Ferrero_TI, Senger_TI} has shown that this design maintains the standard LGAD timing capabilities and that the trenches assure pad isolation for irradiation with neutrons up to a fluence of 3.5\nq[15] and gamma up to 10 MRads. TI-LGADs are the natural evolution of the initial UFSD design as they offer a much higher fill factor without degrading any of the other aspects. 
 
 \subsection{LGAD with resistive read-out}
 
 A recent design innovation has been the introduction of resistive read-out in silicon detectors (RSD). The design, initially proposed with AC-coupled read-out~\cite{TREDI2015N} (RSD or AC-LGAD), has recently been extended to DC-coupled read-out~\cite{DCRSD} (DC-RSD). The sketches of RSD and DC-RSD are shown in Figure~\ref{fig:rsd} where AC signals are shown in red and DC signals in green. Both designs are based on an n-in-p sensor, have a continuous gain implant just underneath the cathode, and the cathode is resistive to ensure pads isolation and signal sharing. 
 
\begin{figure}[h]
\begin{center}
\includegraphics[width=0.5\textwidth]{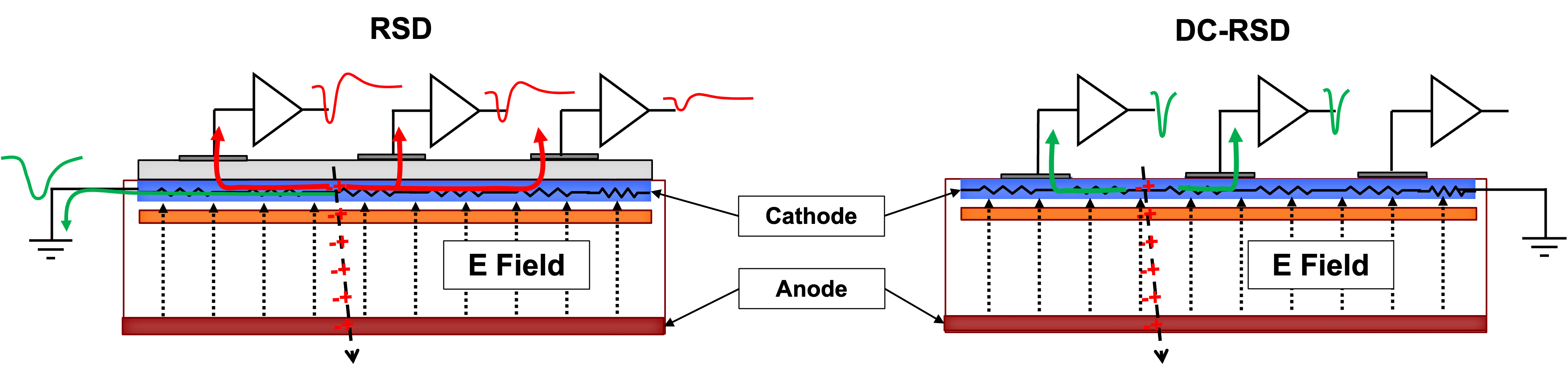}
\caption{Sketch of a resistive silicon detector with AC- or DC- read-out. AC signals are shown in red, while DC signals in green}. 
\label{fig:rsd}
\end{center}
\end{figure}

Signal sharing is built-in in the design, and it functions analogously to a current divider~\cite{tornago2020resistive}: each pad $i$ sees a fraction $I_i$ of the total signal $I_o$ that depends on the impedance $Z_j$ between the impact point and the pads. 

\begin{equation}
\label{eq:divider}
 I_i = I_0 \frac{\frac{1}{Z_i}}{\sum_1^n \frac{1}{Z_j}}.
\end{equation}

RSDs have been tested extensively~\cite{tornago2020resistive, 8846722, apresyan2020measurements,Heller:2022aug}, demonstrating an exceptional position resolution (a few \% of the pitch size) while maintaining the temporal resolution typical of the LGAD design. The major drawback of the RSD design is the difficulty of limiting signal sharing to the closest set of pads: in the present design, the signal remains visible in all the pads located in a radius of about 500-1000 \microns from the hit position. This fact not only complicates the reconstruction but limits the use of RSD in environments with low particle density. Signal sharing can be modeled only for simple pad geometries, so analytic reconstruction methods are mostly not applicable. One approach that has given very good results is the use of machine learning techniques in the reconstruction. %where the signals on the pads are the input data.  
This method is very powerful as it makes use of all aspects of signal propagation and sharing (delays, spreads, relative amplitudes) to identify the most probable hit time and position~\cite{Siviero_2021, SivieroVCI2022}.

In RSD2, the second FBK RSD production~\cite{RSD2}, specially designed electrodes have been introduced to limit signal sharing. First results on structures with cross-shaped electrodes and a pitch of 450 \microns and 1300 \micron~\cite{CartigliaRSD2} showed that for large pitch size, the sharing is contained to a few pixels and that the spatial resolution is excellent, the structure with a pitch of 450 \microns (1300 \micron) has a resolution of $\sigma_x \sim$ 15 (37) \micron. According to simulations, signal sharing in DC-RSD is always limited to a constant number of read-out pads~\cite{MenzioVCI2022} thanks to the introduction of a resistive grid that connects all read-out pads, and that delimits the area of signal sharing within a single grid cell.  

A critical aspect of the RSD design is the possibility of achieving excellent spatial and temporal resolutions with large pixels and thin sensors. 
Sensors need to be thin in experiments that require a very low material budget, such as those at future $e^+e^-$ machines. Therefore signal sharing cannot be obtained using an external magnetic field. In this condition, the only option to achieve a position resolution of 5-10 \microns is to use tiny pixels, about 25x25 \micronsq. However, the power consumption of so many channels is too large, preventing air-cooling use (the present simulation places the limit of air-cooling capabilities at about 0.1 W/cm$^2$). Given the low particle density typical of leptons colliders, large pixels can be used: with RSD sensors, a 5-10 \microns position resolution can be achieved with pixels of about 250 \microns pitch. Figure~\ref{fig:StandardRSD} shows on the left a standard silicon detector while on the right an RSD with the same spatial resolution of about 5-10 \micron. The RSD design has a factor of about 100 fewer pixels, so each read-out amplifier has more space available and can use more power while, at the same time, the total power consumption is considerably lower.

\begin{figure}[h]
\begin{center}
\includegraphics[width=0.5\textwidth]{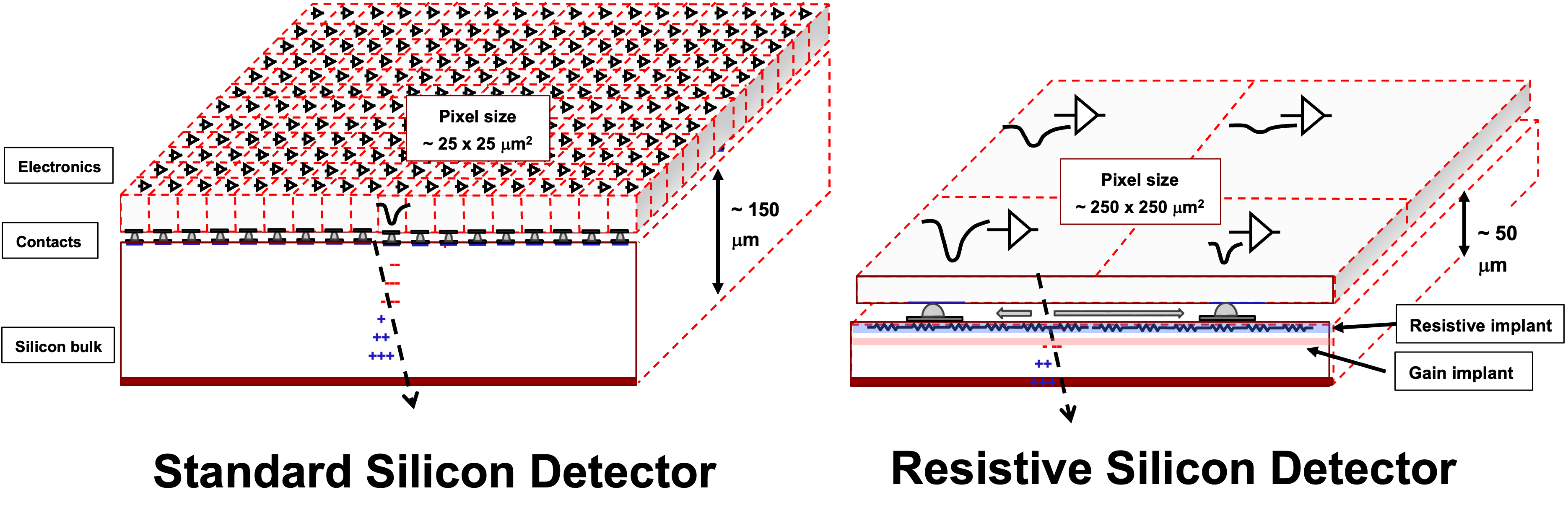}
\caption{ Sketches of a standard silicon detector and of an RSD with the same spatial resolution of about 5-10 \micron.}. 
\label{fig:StandardRSD}
\end{center}
\end{figure}

\subsection{LGAD radiation hardness}
In the standard LGAD design, the gain implant is doped with acceptors (either boron or gallium) that are susceptible to being de-activated by hadron irradiation~\cite{Kramberger_2015, MMoll-Vertex2019}. In this process, called "acceptor removal", the incoming radiation removes a fixed number of acceptors per unit of volume, so it is more damaging for low doped gain implants since, in relative terms, a higher fraction of dopant is removed. It has been demonstrated that the infusion of carbon in the gain layer reduces the acceptor removal rate~\cite{FERRERO201916}: state-of-the-art carbon-infused LGAD sensors can maintain a value of gain of about 10-15 up to fluences of about 2\nq[15]. Such values of gain are maintained by increasing the bias voltage up to about 700V for a 50 \microns thick sensor. However, very recent studies ~\cite{HellerTredi22} have shown that when the electric field in a silicon sensor is about 11.5-12 V/\micron, an impinging MIP can trigger an avalanche process that destroys the sensor (SEB, single event burn-out). The SEB mechanism limits the operating bias voltage on LGAD and therefore reduces the possibility of compensating for acceptor removal. 
Recently, two methods have been proposed to reduce acceptor removal~\cite{SolaVCI2022}: (1) carbon shield and (2) doping compensation. The carbon shield technique is based on the assumption that the vacant states responsible for acceptor removal are very mobile and diffuse from the bulk into the gain implant. If this assumption is correct, they can be stopped by a carbon layer implanted underneath the gain implant. The principle of doping compensation is shown in Figure~\ref{fig:comp}: in the standard LGAD design (i), the gain implant is obtained via a single p-implant, and acceptor removal decreases the effective doping (iii). In the compensated design, (ii), the gain implant is obtained as the difference between two implants, a larger p-implant, and a smaller n-implant. (iv) The effect of irradiation is the concurrent reduction of acceptors and donors. The effective gain implant can increase, decrease, or remain constant depending on the presently unknown acceptors and donors removal rates within this type of compensated implant. Both techniques are presently being implemented in the ExFlu1 production at FBK.

\begin{figure}[h]
\begin{center}
\includegraphics[width=0.5\textwidth]{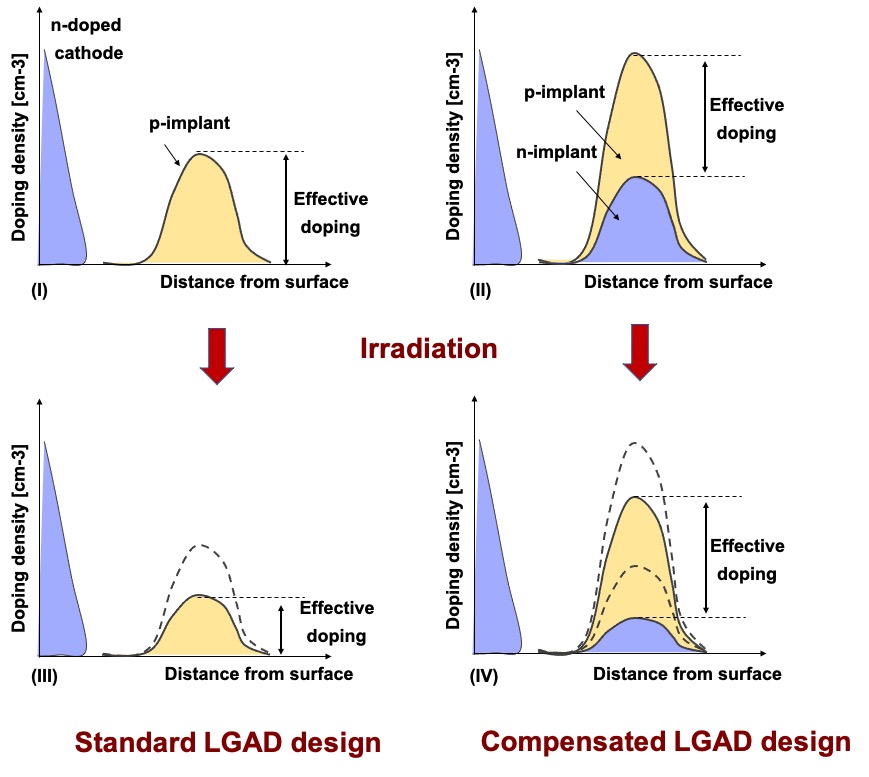}
\caption{ Sketches of the compensation technique. Left: effect of irradiation on a standard LGAD. Right: effect of irradiation in a compensated LGAD}. 
\label{fig:comp}
\end{center}
\end{figure}

\section{Conclusions}
The field of 4D tracking is experiencing a very fast evolution, with advances in MAPS and hybrid systems. Presently, two large LGAD-based timing layers, aiming at a temporal resolution of about 45 ps/hit with a spatial resolution of $\sim$ 375\micron, are under construction, one in the ATLAS experiment and one in the CMS experiment. These timing layers are the first stepping stones toward systems able to perform real 4D tracking. Several small prototypes are under development, using both the hybrid designs (TIMESPOT, TimePix) and MAPS (FASTPIX, MonPicoAD, miniCACTUS). Presently, the most challenging aspect of a 4D tracker is the front-end electronics design due to the limited space and power consumption.

The LGAD design is evolving to overcome its present limitations. The introduction of trenches to separate pixels has reduced by almost ten the inter-pad no-gain distance, from 50-80 \microns to about 5 \micron. Currently, LGADs work with unchanged performances up to a fluence of about 1\nq[15]: two new techniques, carbon shield and compensated gain layer, might considerably extend this value. The introduction of resistive read-out in the LGAD design allows using very large pixels while maintaining excellent temporal and spatial resolutions; this design might considerably reduce power consumption since it uses almost a factor of 100 fewer read-out channels. 

The R\&D activities in the next 5-10 years will be critical in defining the technologies available for the next generation of experiments; the community must find the resources to develop the enabling technologies for 4D tracking. 

\section*{Acknowledgments}

We thank our collaborators within RD50, ATLAS, and CMS, who participated in the development of UFSD. We kindly acknowledge the following funding agencies and collaborations: INFN – FBK agreement on sensor production; Dipartimenti di Eccellenza, Univ. of Torino (ex L. 232/2016, art. 1, cc. 314, 337); Ministero della Ricerca, Italia, PRIN 2017, Grant 2017L2XKTJ – 4DinSiDe; Ministero della Ricerca, Italia, FARE, Grant R165xr8frt\_fare

\bibliography{NC_bibfile_2022}

\end{document}